\documentclass{PoS}
\usepackage{graphicx}

\title{Target characterization for the $^{130}$Ba($\alpha,\gamma$)$^{134}$Ce $\gamma$-process experiment}

\ShortTitle{Target characterization for the $^{130}$Ba($\alpha,\gamma$)$^{134}$Ce $\gamma$-process experiment}

\author{Gy. Gy\"urky, \speaker{Z. Hal\'asz}, J. Farkas, Zs. F\"ul\"op, E. Somorjai and  T. Sz\"ucs\\
        Institute of Nuclear Research (ATOMKI), H-4001 Debrecen, POB.51., Hungary\\
        E-mail: \email{\{gyurky|fiser|jafar|fulop|somorjai|tszucs\}@atomki.hu}}

\abstract{In order to extend the available experimental database for the astrophysical $\gamma$-process, the cross section measurements of the $^{130}$Ba($\alpha,\gamma$)$^{134}$Ce and $^{130}$Ba($\alpha$,n)$^{133}$Ce reactions are in progress. The measurements are carried out using thin layers of evaporated BaCO$_3$ as target. Since the target thickness enters directly into the calculation of the cross sections, the reliability of  its determination is of crucial importance. Three different methods have been used to determine the target thickness. Details of these experiments and the obtained results are presented.}

\FullConference{11th Symposium on Nuclei in the Cosmos, NIC XI\\
		July 19-23, 2010\\
		Heidelberg, Germany}

\begin{document}

\section{Introduction}

The astrophysical $\gamma$-process, as the main part of the more general p-process is responsible for the stellar production of the heavy, proton rich isotopes inaccessible to the neutron capture s- and r-processes \cite{arn03}. Modeling of the $\gamma$-process takes into account a huge reaction network of tens of thousands of reactions with the aim of reproducing the abundances of the produced p-isotopes observed in the Solar System. $\gamma$-induced reactions play the main role in the reaction networks, but other reactions, such as radiative captures are also included. 

Due to the large number of reactions involved and the lack of experimental data the $\gamma$-process networks utilize theoretical reaction rates obtained typically form cross section of the Hauser-Feshbach statistical model calculation. Any uncertainty in the calculated cross sections will influence the resulting p-isotope abundances. The relatively poor predictive power of modern $\gamma$-process models can at least in part be attributed to incorrect reaction rates \cite{dil08}. Therefore, the measurement of reaction cross sections relevant for the $\gamma$-process is important to make the model calculations more reliable.

It has been shown that the calculations are especially sensitive to the rates of reactions involving $\alpha$-particles \cite{rau06,rap06}. Moreover, these are the reactions where experimental data are almost completely missing \cite{gyu10}. Therefore, increasing experimental effort is devoted to the study of $\alpha$-induced reactions. The next step in this systematic study is the measurement of the $^{130}$Ba($\alpha,\gamma$)$^{134}$Ce and $^{130}$Ba($\alpha$,n)$^{133}$Ce reaction cross sections which is currently under way at the Institute of Nuclear Research (ATOMKI) in Debrecen, Hungary. In the present paper one important aspect of the experimental procedure, the determination of the Ba target thickness is detailed.  

\section{Preparation and characterization of the Ba targets}

The abundance of the $^{130}$Ba isotope in natural Barium is very low, only 0.1\,\%. Therefore, in order to measure the relatively low cross sections of the  $^{130}$Ba($\alpha,\gamma$)$^{134}$Ce and $^{130}$Ba($\alpha$,n)$^{133}$Ce reactions, isotopically enriched $^{130}$Ba targets have to be used. Suppliers provide this isotope in carbonate (BaCO$_3$) form, this compound has been used as the starting material for target preparation. The targets have been prepared by vacuum evaporation onto thin (2\,$\mu$m) Al foil backings. $^{130}$BaCO$_3$ powder has been placed into a Ta crucible heated by AC current. The Al foil backings were positioned about 5 cm above the crucible. So far nine targets have been prepared and used for cross section measurements.

It is known that BaCO$_3$, like other alkaline-earth carbonates may undergo thermal decomposition when heated to high temperatures (see e.g. \cite{lvo04} and references therein). Therefore the chemical composition of the evaporated targets needs careful investigation. Since the important quantity for the cross section determination is the areal number density of the $^{130}$Ba atoms, after determining the stoichiometry of the targets, two further independent methods have been used to determine this quantity. In the following sections the details of these measurements are given.

\subsection{Target characterization by Rutherford Backscattering Spectrometry}

\begin{figure}
\includegraphics[angle=-90,width=0.7\textwidth]{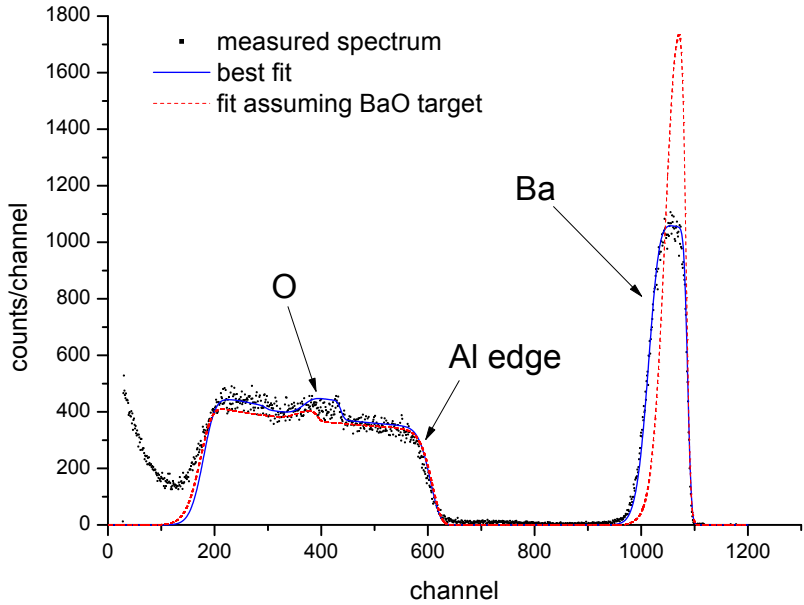}
\caption{\label{fig:RBS}RBS spectrum of target \#6. See text for details.}
\end{figure}

The composition of the targets and the number density of the $^{130}$Ba atoms have been measured with the Rutherford Backscattering Spectrometry (RBS) technique. The Van de Graaff accelerator of ATOMKI provided a 2.5\,MeV $\alpha$-beam which was scanned over the target surface of about 8\,mm in diameter. A collimated Si detector was built into the target chamber at 150$^\circ$ with respect to the beam direction to detect the backscattered $\alpha$ particles. The measured RBS spectra have been fitted with the SIMNRA code \cite{SIMNRA}. The number density of the Ba atoms and the target stoichiometry were free parameters in the fit. The best fit was obtained with the following stoichiometry: Ba : C : O = 19\,\% : 20\,\% : 61\,\%. Since the fit is only weakly sensitive on the Carbon content of the target, the C ratio has been fixed at 20\,\%. These results show that there is no significant deviation from the stoichiometric BaCO$_3$ composition of the target after the evaporation. Figure \ref{fig:RBS} shows the RBS spectrum of target \#6 and the fit to the data. For comparison, a fit assuming BaO target composition (with the assumption that BaCO$_3$ completely loses CO$_2$ and forms BaO during the evaporation) is shown. The fit was made to obtain the same fitted and measured Ba peak area. It can clearly be seen  that the BaO target composition can be excluded.

\subsection{Independent determination of the target thickness}

Relying on the target stoichiometry obtained from the RBS measurement, the number of target atoms has also been determined by two other independent methods. First, the weight of the Al foil backing was measured with a precision of a few $\mu$g before the evaporation. The foils were put into the evaporator in a holder constraining the condensation of the target to a circular spot of 12\,mm in diameter. The weight of the target was measured again after the evaporation and from the weight difference the target thickness could be obtained. This method assumes a uniform target thickness across the 12\,mm diameter surface.

\begin{figure}
\includegraphics[angle=-90,width=0.7\textwidth]{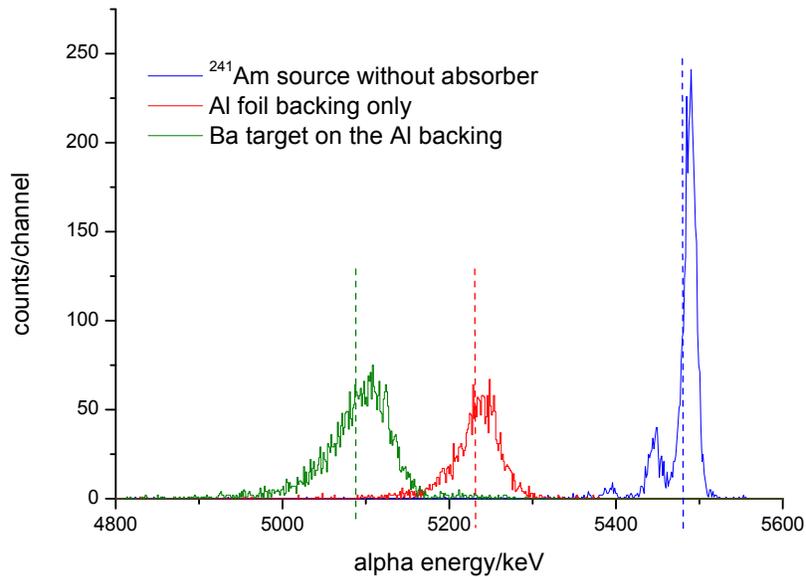}
\caption{\label{fig:energyloss}Alpha spectrum of the $^{241}$Am source without any absorber, spectrum after energy loss in the Al backing and in the Ba target.  The center of gravity of each spectra are shown by dashed lines.}
\end{figure}

In the second method the energy loss of $\alpha$-particles in the target was used to determine the thickness. A $^{241}$Am $\alpha$-source and a Si detector have been placed into an ORTEC SOLOIST $\alpha$-spectrometer. First the spectrum of the source was measured. Then the Al backing (before the evaporation) has been put between the source and detector and the spectrum was measured again. Finally after the evaporation the target was placed in the spectrometer to measure the $\alpha$-particles penetrating the target and backing. The three spectra for target \#2 can be seen in Figure\,\ref{fig:energyloss}. $^{241}$Am emits three main $\alpha$-groups which can be separated in the case of the naked $^{241}$Am spectrum. After the target or backing, however, the energy straggling makes it impossible to separate the subgroups. Therefore the energy loss was determined by calculating the center of gravity of the spectra in all three cases. These centers can be seen as vertical dashed lines in the figure. The energy loss was converted into target thickness using the SRIM code \cite{SRIM}. 

The latter method has also been used to determine the uniformity of the target thickness. For this purpose a 2\,mm collimator has been placed between the source and detector and moved over the surface of the target. The uniformity of the targets have been found to be better than 7\,\%.  

\section{Results and conclusion}

\begin{figure}
\includegraphics[angle=-90,width=0.7\textwidth]{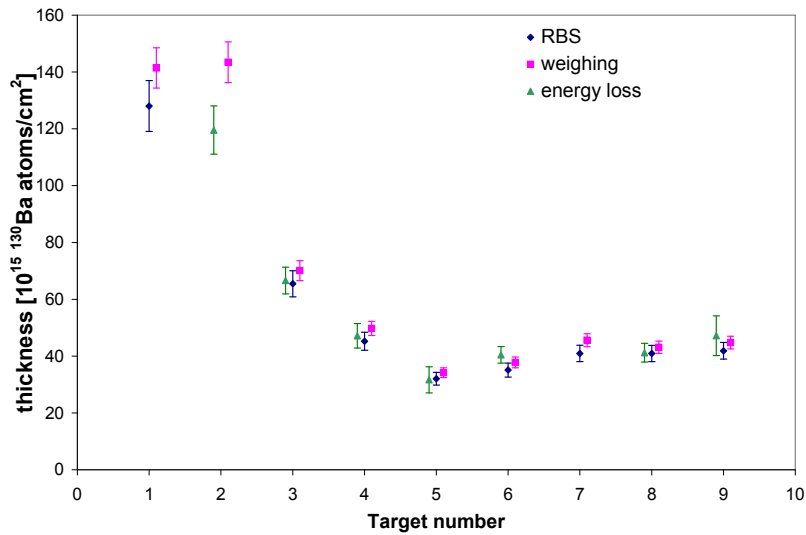}
\caption{\label{fig:results} Thicknesses of the Ba targets determined by the three methods.}
\end{figure}

The results of the three independent thickness measurements can be seen in Figure\,\ref{fig:results}. In a few cases not all three methods have been applied because of technical reasons. The results of the different methods are in excellent agreement with exception of target \#2 where the results of the two applied methods differ by more than the uncertainties. In order to avoid higher systematic uncertainty, the cross section measurement carried out with this target will be repeated with new targets. 

The typical uncertainties for the thickness determinations are the following: 7\,\% for the RBS, 5\,\% for the weighing and 7\,\% to 14\,\% for the alpha energy loss measurement. The latter two are, however, not entirely independent from the RBS measurement, since the target stoichiometry must be used as an input parameter. Therefore the unweighted average of the three methods has been adopted as the final result of the target thickness and a conservative value of 8\,\% has been assigned as uncertainty. 

The three applied methods increase the robustness of the target thickness determination. Nevertheless, the number of target atoms remains one of the most significant source of uncertainty in the $^{130}$Ba($\alpha,\gamma$)$^{134}$Ce and $^{130}$Ba($\alpha$,n)$^{133}$Ce cross section measurement.
Further investigations of the target properties are planned with SNMS \cite{vad09} and SEM \cite{cse03} techniques.

\acknowledgments

This work was supported by the European Research Council grant
agreement no. 203175 and OTKA grants K68801 and NN83261(Eurogenesis).

\end{document}